\documentclass[aps, prd, twocolumn, lengthcheck, superscriptaddress, 
nofootinbib]{revtex4-1}

\usepackage{epsfig}
\usepackage[usenames]{color}
\usepackage{graphicx}
\usepackage{amsmath}
\usepackage{epstopdf}

\def\bi{\bibitem}

\def\la{\langle}
\def\ra{\rangle}
\def\be{\begin{eqnarray}}\def\ee{\end{eqnarray}}
\def\lsim{\mathrel{\rlap{\lower3pt\hbox{\hskip1pt$\sim$}}
     \raise1pt\hbox{$<$}}} 
\def\gsim{\mathrel{\rlap{\lower3pt\hbox{\hskip1pt$\sim$}}
     \raise1pt\hbox{$>$}}} 
\def\del{\partial}

\allowdisplaybreaks

\def\Tr{\rm Tr}
\def\bi{\bibitem}

\newcommand{\CL}{{\cal L}}

\newcommand{\Slash}[1]{\ooalign{\hfil/\hfil\crcr$#1$}}

\begin{document}

\title{ Dichotomy of Baryons as Quantum Hall Droplets and Skyrmions:\\
Topological Structure of  Dense Matter}

\author{Yong-Liang Ma}
\email{ylma@ucas.ac.cn}
\affiliation{School of Fundamental Physics and Mathematical Sciences,
Hangzhou Institute for Advanced Study, UCAS, Hangzhou, 310024, China}
\affiliation{International Center for Theoretical Physics Asia-Pacific (ICTP-AP) (Beijing/Hangzhou), UCAS, Beijing 100190, China}

\author{Mannque Rho}
\email{mannque.rho@ipht.fr}
\affiliation{Universit\'e Paris-Saclay, CNRS, CEA, Institut de Physique Th\'eorique, 91191, Gif-sur-Yvette, France }

\date{\today}

\begin{abstract}

We review the recent exploration of a possible ``domain-wall structure" of compressed baryonic matter  in massive compact stars in terms of fractional quantum Hall droplets and skyrmions for baryons in medium. The theoretical framework is anchored on an effective nuclear effective field theory that incorporates two hidden symmetries, flavor local symmetry  and scale symmetry conjectured to be dual to the gluons and quarks of QCD.   It hints at a basically different, hitherto undiscovered structure of nuclear matter at low as well as high densities. Hidden ``genuine dilaton (GD)" symmetry and hidden local symmetry (HLS) gauge-equivalent at low density to nonlinear sigma model capturing chiral symmetry,  put together in nuclear effective field theory, are seen to play an increasingly important role in providing  hadron-quark duality in  baryonic matter. This strongly motivates incorporating both symmetries in formulating  ``first-principles" approaches  to nuclear dynamics encompassing from the nuclear matter density to the highest density stable in the Universe.

\end{abstract}
\maketitle
\setcounter{footnote}{0}

\section{Introduction}
In this  review note, we  recount what we have done in the past few years to uncover a totally novel structure, not found in the literature, of dense nuclear matter relevant to massive compact stars. Our approach is anchored on two symmetries hidden in dilute hadronic systems, i.e., chiral symmetry and scale symmetry,  that could play a crucial role as density increases high. In the accompanying article written by one of us (MR), how the very symmetries involved in dense compact-star matter manifest also at nuclear matter density in one of the outstanding problems in nuclear physics lasting more than four decades, namely, the ``quenching" of the axial-vector coupling constant  $g_A$ in nuclear Gamow-Teller transitions.

The structure of highly dense  matter found in massive compact stars is a totally uncharted domain. Unlike at high temperature,  at high density, it can be accessed neither by lattice QCD nor by  terrestrial experiments. While,  as comprehensively reviewed recently~\cite{vankolck,holt-rho-weise}, finite nuclei as well as  infinite nuclear matter can be fairly accurately accessed by nuclear effective field theories,  pionless or pionful, referred to herewith as ``standard nuclear effective field theory ($Sn$EFT)"  anchored on relevant symmetries and invariances along the line of Weinberg's Folk Theorem~\cite{folktheorem}, $Sn$EFTs, as befits their premise, are expected to break down at some high density (and low temperature) relevant to, say, the interior of massive stars.

In $Sn$EFT, the power counting in density is $O(k_F^q)$ where $k_F$ is the Fermi momentum and increasing density involves going to higher $q$. For the ``normal" nuclear matter with density $n_0\approx 0.16$ fm$^{-3}$, the expansion is required to go up to $q\sim 5$~\cite{holt-rho-weise}. The breakdown must occur as $k_F$ goes beyond the nuclear matter density. On the other hand, if one organizes  effective field theories  in renormalization-group approach built on a Fermi surface, the power counting in $k_F$ comes out to be $O((1/\bar{N})^\kappa)$ where $\bar{N}=k_F/(\tilde{\Lambda} -k_F)$  with $\tilde{\Lambda}$ being the cutoff on top of the Fermi sea measured relative to the center of the Fermi sphere.  The expansion in $\kappa \geq 0$  leads to Fermi-liquid fixed-point theory~\cite{shankar,polchinski}.  The equilibrium nuclear matter is given at the Fermi-liquid fixed point with $\bar{N}\to \infty$. Approaching Fermi-liquid theory starting from $Sn$EFT for nuclear (or neutron) matter valid up to roughly  $\sim n_0$ has been formulated~\cite{holt-kaiser-weise, holt-rho-weise}. It has also been formulated using the $V_{lowk}$ renormalization-group (RG) approach applicable to both finite nuclei and infinite matter taking into account $1/\bar{N}$ corrections~\cite{brown-kuo}. 

Given that the $k_F$ expansion must inevitably break down -- and hence $Sn$EFT becomes no longer valid -- at some high density above $n_0$,   a potentially promising and justifiable approach is to go over to the Fermi-liquid structure starting from the normal nuclear matter density at which the Fermi-liquid structure is fairly well established  to hold. Our strategy is to build a model, that we shall refer to as ``G$n$EFT," that while capturing fully what $Sn$EFT successfully does up to $n_0$,  can be extrapolated up beyond the density at which  $Sn$EFT is presumed  to break down. Such an approach developed in \cite{PKLMR,MR-review} is  anchored on a Lagrangian that incorporates, in addition to the pions and nucleons of , the lowest-lying vector mesons $\rho$ and $\omega$  and the scalar meson $\chi$ standing for $f_0 (500)$.   We treat the vector mesons $V=(\rho,\omega)$ as dynamical fields of hidden local symmetry (HLS)~\cite{HLS84} -- equivalently ``composite gauge fields"~\cite{suzuki} -- and the scalar $\chi$ as a ``genuine dilaton (GD)"~\cite{crewther},   a (pseudo-)Nambu-Goldstone (NG) boson of hidden scale symmetry~\cite{CT}.  
We cannot say whether these symmetries are intrinsic in QCD. If they are intrinsic then they must be hidden in the vacuum since they are not visible. What happens in our systems in nuclear medium is that they get ``unhidden" by, or emerge from, strong nuclear correlations as nuclear matter is highly compressed. One of our basic assumptions is that the HLS is consistent with the Suzuki theorem~\cite{suzuki} and the scale symmetry with genuine dilaton has an infrared (IR) fixed point with both the chiral and scale symmetries realized in the NG mode~\cite{crewther} at some high density.  How these symmetries, invisible in free space, could appear in dense medium has been the subject of the past efforts~\cite{PKLMR,MR-review}  in nuclear astrophysics and motivates going beyond what has been explored so far. We approach this issue by analyzing  the structure of cold dense baryonic matter with density $n > (2-3)n_0$ in terms of a ``baryon-quark duality"\footnote{Note that the accent, as is to be clarified, is on ``duality," not on ``continuity" discussed in connection with confinement-deconfinement issue.} in QCD.

In this paper we find that the combined hidden scale symmetry and HLS, suitably formulated so as to access high density compact-star matter as in \cite{PKLMR,MR-review}, interpreted as ``emergent" from strong nuclear correlations, reveals a dichotomy in the structure of baryons treated in terms of topology in the large $N_c$ approximation and discuss how it could affect the equation of state (EoS) at high density relevant to the cores of massive compact stars. The merit of this work is that it exploits in strongly-interacting baryonic matter a certain ubiquitous topological structure of highly correlated fermions like electrons in condensed matter, thereby bringing in a possible paradigm change in nuclear dynamics.

We should admit that given the highly exploratory nature of our work drawing from what we have learned of compact-star matter which is distinctively different from standard approaches found in the literature -- and the paucity of reliable model-independent approaches to dense baryonic matter, some of the arguments we develop are inevitably speculative or conjectural.

\section{The Problem: Dichotomy}
\label{problem}
Consider the baryons made up of the quarks with nearly zero current quark masses. We will be dealing primarily with two flavors, u(p) and d(own). However for the role of scale symmetry, it is essential to think in terms of 3 flavors~\cite{crewther} as we will explain below.  For three flavors,  all octet baryons $B^{(\alpha)}$, $\alpha=1, ..., 8$,  can be obtained as solitons, i.e., skyrmions~\cite{skyrme},  from the octet mesons. This has to do with the homotopy group $\pi_3 (SU(3)) \simeq \cal Z$ and is justified in QCD at the large $N_c$ limit. However there has been one annoying puzzle in this matter:  There is no skyrmion associated with  the singlet meson $\eta^\prime$. This is because $\pi_3 (U(1))=0$. The resolution to this conundrum was suggested in 2018 by Komargodski~\cite{zohar}: The baryon for $\eta^\prime$, while not a skyrmion soliton, turns out to be also a topological object at the large $N_c$ limit but more appropriately a fractional quantum Hall (FQH) droplet,  somewhat like a pancake (or perhaps pita~\cite{karasik,karasik2}).

One way of seeing how this FQH droplet comes about in QCD, the approach we adopt in this paper,  is in terms of the ``Cheshire Cat phenomenon" (CCP)  formulated a long time ago~\cite{CCP}. 

In the CCP, the trade-in of topology for hadron-quark continuity for low energy/long wavelength nuclear processes involving the u quark and d quark  is via the ``infinite hotel mechanism" where  $N_c$ quarks disappear into the vacuum, with the baryon charges taken up by the triplet pions, i.e., skyrmions . The $N_f=1$ quarks, on the other hand, forbidden to fall into the infinite hotel and become skyrmions,  go instead into a 2-dimensional quantum Hall (QH) droplet described by Chern-Simons (CS) theory. In  \cite{MNRZ} this was described as the anomaly cancelation,  known as the ``anomaly inflow" from the bulk of a system to its boundary~\cite{callan-harvey}. Here what's involved is that the quarks propagating in one direction flow to one higher dimension making a sheet in $(x, y)$ with the anomaly caused by the ``confinement" boundary condition,  with the resulting system given by abelian CS theory for the FQH droplet. One can think of this as a topological object of $\eta^\prime$ with a topology  different from that of the skyrmions of $\pi$'s. We denote this baryon $B^{(0)}$. 

There is another way of interpreting the CCP construction of the fractional quantum Hall droplet that could be more appropriate for resolving the ``dichotomy problem" mentioned below. That is to formulate it in terms of a domain-wall (DW) structure.  Consider the confinement wall at $x_3=0$. The vacua at $x_3>0$ and $x_3<0$ are clearly different. Then the confinement wall makes the DW at $x_3=0$. As an example,  think of the region $x_3 <0$ as the region in which quarks are confined in the sense of the MIT bag. Then calculating the spectral asymmetry in the limit of the thin wall with the chiral bag boundary condition set at $x_3=0$, one reproduces the same baryon charge obtained in the anomaly-inflow mechanism~\cite{mapping}. 

One important consequence of this observation is that the QH droplet has the spin $J=N_c/2$, namely 3/2 for $N_c=3$, corresponding to the baryon resonance $\Delta (3/2,3/2)$.  The mass of the $B^{(0)}$  in the large $N_c$ limit is of course $\sim O(N_c)$ but it can also receive $O(N_c^0)$ contribution~\cite{zohar}. In the skyrmion system, there is also a baryon of the same quantum numbers  $(3/2,3/2)$ but there is no correction coming at $O(N_c^0)$  that distinguishes spin-1/2 and spin-3/2.\footnote{There is the Casimir contribution to the skyrmion mass that comes at $O(N_c^0)$ but that is common to the skyrmions of both spins 1/2 and 3/2.}  The first correction to the skyrmion mass in the large $N_c$ limit comes at $O(1/N_c)$ arising from the rotational quantization.  This presents a ``dichotomy problem" (dubbed DP for short).

One can see this dichotomy if one  applies  the same argument made for the CCP for the QH droplet for $B^{(0)}$ to $N_f=2$ systems, namely the nucleon. Instead of dropping into the ``infinite hotel" in the CC mechanism for the $N_f=2$ skyrmion when the bag is shrunk to zero~\cite{CCP,RGB,GJ}, there seems to be nothing that would prevent the  quarks from undergoing the anomaly inflow into fractional quantum Hall droplets making the CS theory nonabelian~\cite{MNRZ}. Why not form a sheet-structured matter arranged, say, in the lasagne arrays seen at high density in crystal lattice simulations of compact-star matter (to be mentioned below)?

The pertinent question then is: What dictates the $N_f=2$ quarks to (A) drop in the $\infty$-hotel skyrmions or (B) instead to  flow to nonabelian FQH droplets? Or could it be into (A) {\it and} (B) in some combination? This sharpens and generalizes the dichotomy problem raised above. A solution to this dichotomy problem is recently addressed by Karasik~\cite{karasik} in terms of a ``generalized" current that {\it unifies} the $N_f=1$ baryon,  QH droplet,   and the $N\geq 2$ baryon,  skyrmion. Here we explore whether and how  one can go from one to the other for the $N_f=2$ systems in terms of the EoS for dense baryonic matter. We do this by ``dialing" baryon density. The hope  is that this will unravel the putative {\it hadron-quark  duality} possibly involved in the physics of massive compact stars. The strategy here is to extract the conceptual insights gained in the phenomenological development discussed in \cite{MR-review} for the physics of massive compact stars, the only system currently available in nature for high density $n \gg n_0$ at low temperature and  translate them into a scheme that could address, at least qualitatively, the dichotomy problem.
\section{G$n$EFT Lagrangian}
We begin by writing the effective Lagrangian involved, in as simple a form as possible, that allows us to capture the basic idea developed. The details look rather involved, but  the basic idea is in fact quite simple. We will first deal with the mesonic sector with baryons generated as solitons and later explicitly incorporate baryons. In developing the basic idea, we will frequently switch back and forth between the former description and the latter.
\subsection{Scale-invariant hidden local symmetric (sHLS) Lagrangian}
To address the dichotomy problem (DP) in  highly compressed baryonic matter,  we  incorporate  the $\eta^\prime$ field, $\eta^\prime\in U_A(1)$, in addition to the pseudo-scalar NG bosons $\pi\in SU(2)$ and the vectors $\rho_\mu \in SU(2)$ and $\omega\in U(1)$.  The reason for this, as we will argue, is that although $\eta^\prime$ is massive compared with the mesons we will take into account, it  goes massless in the limit that $N_c\to\infty$ and plays a crucial role for bringing in Chern-Simons topological field theory structure in (possibly) dense baryonic systems. In the modern development, it is being suggested that the role of $\eta^\prime$ in the guise of fractional quantum Hall ``pancake" for the flavor singlet baryon $B^{(0)}$ plays an indispensable role for QCD phase structure at high density. We will comment on this matter later although it is not well understood at present.

The Crewther's GD approach to scale symmetry~\cite{crewther} adopted in this paper~\cite{MR-review} -- which will turn out to play a crucial role in our theory -- necessitates the kaons on par with the dilaton. For our purpose, however, we can ignore the strange quark -- given the presence of the $\eta^\prime$ meson -- and focus on the two light flavors.  For the reason that will become clear later, unless otherwise specified, the $\rho$ and $\omega$ fields will  be treated in $U(2)$-nonsymmetric way.\footnote{In the Seiberg-dual approach to HLS, the $\omega$ meson  is not pure $U(1)$ of $U(N_f)$ that contains $\rho$ but a mixture of $U(1)$s. This feature will appear later in  Sect.~\ref{withetaprime} where baryonic matter with $\eta^\prime$ present is treated.}\label{ZZ}

We write the chiral field  $U$ as~\footnote{{In this paper, we use the convention of \cite{HY:PR}}}
\be
U= \xi^2=e^{i {\eta^\prime}/{f_\eta}} e^{i\tau_a\pi_a/f}
\ee
and the HLS fields as
\be
V_\mu^\rho& = &{} \frac{1}{2}g_\rho \rho_\mu^a\tau^a, \quad V_\mu^\omega = {} \frac{1}{2}g_\omega \omega_\mu.
\ee
Expressed  terms of the Maurer-Cartan 1-forms
\be
\hat{\alpha}_{\parallel,\perp}^\mu & = & \frac{1}{2i}\left(D^\mu \xi \cdot \xi^\dagger \pm D^\mu \xi^\dagger\cdot \xi\right)
\ee
where $D_\mu \xi = (\partial_\mu - iV_\mu^\rho - i V_\mu^\omega)\xi$, the HLS Lagrangian we are concerned with is of the same form as the HLS Lagrangian for 3 flavors~\cite{HY:PR} with the parity-anomalous homogeneous Wess-Zumino (hWZ)  Lagrangian composed of 3 terms (in the absence of external fields). For the $SU(2)\times U(1)$ case we are dealing with, there is no five-dimensional Wess-Zumino (WZ) term.

To implement scale symmetry \`a la GD~\cite{crewther},  we are to use as explained below the conformal compensator field $\chi=f_\chi e^{\chi/f_\chi}$ which has both mass dimension and scale dimension 1. The structure of the Lagrangian that we use in this discussion was written down before in our reviews~\cite{MR-review} and we shall write it down here again with $\eta^\prime$ incorporated. In oder to justify its extension in the density domain where FQH droplets could play a role, an additional ingredient to what figures in \cite{MR-review} is needed. It will be given below.

The scale-symmetrized Lagrangian that we denote as sHLS is of the form
\be
{\CL}_{sHLS}
& = & f^2\Phi^2\text{Tr}\left(\hat{\alpha}^\mu_\perp\hat{\alpha}_{\perp\mu}\right) + f_{\sigma\rho}^2\Phi^2\text{Tr}\left(\hat{\alpha}^\mu_\parallel\hat{\alpha}_{\parallel\mu}\right)\nonumber\\
& &{} + f_{0}^2\Phi^2\text{Tr}\left(\hat{\alpha}^\mu_\parallel\right)\text{Tr}\left(\hat{\alpha}_{\parallel\mu}\right) \ + {\CL}_{hWZ\nonumber}\\
& &{} - \frac{1}{2g_\rho^2}\text{Tr}(V_{\mu\nu}V^{\mu\nu}) - \frac{1}{2g_0^2}\text{Tr}(V_{\mu\nu})\text{Tr}(V^{\mu\nu}) \nonumber \\
& &{} +\frac{1}{2}\partial_\mu \chi \partial^\mu \chi + V(\chi)
\label{eq:LOLL}
\ee
where ${\CL}_{hWZ}$  is the hWZ term that conserves parity and charge conjugation but violates intrinsic parity, $\Phi$ is defined as
\be
\Phi = \chi/f_\chi
\ee
and $V_\chi$ is the dilaton potential, the explicit form of which is not needed for our purpose. $V^{\mu\nu}$ is the usual field tensor
\be
V^{\mu\nu} & = & \partial^\mu V^\nu - \partial^\nu V^\mu - i [V^\mu, V^\nu]
\ee
with $V^\mu = V_\rho^\mu + V_\omega^\mu$.
In \eqref{eq:LOLL}
\be
f_0^2 & = & \frac{f_{\sigma\omega}^2 - f_{\sigma\rho}^2}{2}, \qquad \frac{1}{g_0^2} = \frac{1}{2}\left(\frac{1}{g_\omega^2} - \frac{1}{g_\rho^2}\right),
\ee
where $f_{\sigma V}$ figures in the mass formula $m_V^2=f^2_{\sigma V}g_V^2$\footnote{The familiar example is for {$V_\mu =\frac{1}{2}g_\rho \rho_\mu^a\tau^a$}, $f^2_{\sigma\rho}=2f_\pi^2$, giving the KSRF relation, $m_\rho^2=2 f_\pi^2 g_\rho^2$.}.  The $U(2)$ symmetry is recovered at the classical level by setting $g_\omega=g_\rho$ and {$f_0=1/g_0=0$} in (\ref{eq:LOLL}).  The hWZ Lagrangian that will be found to unify the $N_f=1$  and $N_f\geq 2$ baryons consists of three terms (ignoring the external fields not needed in our case)
\be
{\CL}_{hWZ}= \frac{N_c}{16\pi^2}\sum _{i=1}^{3} c_i {\CL}_i\label{hWZterm}
\ee
with
\be
{\CL}_1 &=& i{\Tr} [\hat{\alpha}_L^3  \hat{\alpha}_R - \hat{\alpha}_R^3 \hat{\alpha}_L], \nonumber\\
{\CL}_2 &=& i{\Tr} [\hat{\alpha}_L \hat{\alpha}_R\hat{\alpha}_L \hat{\alpha}_R], \nonumber\\
{\CL}_3 &=& i{\Tr} V [\hat{\alpha}_L \hat{\alpha}_R - \hat{\alpha}_R \hat{\alpha}_L] .
\ee

We should make two remarks on the Lagrangian (\ref{eq:LOLL}) that need to be kept in mind in what follows. First it is $O(p^2)$ in power counting~\cite{HY:PR}  except for the hWZ term which while quartic is indispensable for unifying FQH droplets and skyrmions~\cite{karasik}. Second it is made scale-invariant by the conformal compensator except for the dilaton potential $V(\chi)$ which could contain scale-symmetry explicit breaking, e.g., quark mass terms. The rationale for this strategy for scale symmetry is explained below.
\subsection{``Genuine dilaton" scenario (GDS)}
In accordance with the GD scheme~\cite{crewther} with the IR fixed point specified above, even if explicit symmetry breaking is ignored, scale symmetry can be spontaneously broken by dilaton condensate generating masses to the hadrons. The scheme follows roughly the line of ideas based on ``hidden quantum scale invariance"~\cite{QSI}.  The underlying reason is that sHLS that we are exploiting is connected with strong-weak dualities \`a la Seiberg, typically associated with supersymmetric gauge theories~\cite{Y,Komargodski,abel-bernard,kitanoetal}. This connection allows us to exploit the possible duality of HLS to the gluons that will figure in the problem. In addition, the applicability of duality in non-supersymmetric case as ours is made feasible if scale symmetry is broken by the dilaton in terms of the conformal compensator~\cite{abel-bernard}.

From the point of view of our bottom-up approach, it is important to note that the HLS we are dealing with is dynamically generated~\cite{HLS84}.  This means that the coefficients $c_i$s in the hWZ term that play a key role in the unification of solitonic description of both $N_f=1$ and $N_f\geq 2$ baryons are constants that cannot be fixed by the theory. For low density, therefore, they are to be determined by experiments\footnote{They could be fixed by holographic QCD, but there are no known holographic QCD models that possess possible ``orange"~\cite{karasik}-- not to mention ultraviolet --  completion that would allow approaching the density.}. However as we will see later, in approaching QH droplet baryons bottom-up in density, the coefficients will be ``quantized" by topology~\cite{karasik,karasik2}.  This takes place because there can be a phase transition from a Higgs mode  to a topological phase in which the HLS fields are supposed to be (Seiberg-)dual to the gluons of QCD~\cite{Y}. 
\subsubsection*{Unhiding hidden scale symmetry in nuclei}
Underlying the idea of duality in hadronic matter is that scale symmetry is hidden. The VeV of the dilaton $\chi$ breaks scale-symmetry spontaneously in nuclear medium and slides with the density of the matter. This spontaneous scale-symmetry breaking in nuclear medium via the dilaton condensate dependent  on density  can have a highly subtle impact on certain nuclear properties. One of the celebrated cases highlighting the possible restoration of scale symmetry by nuclear renormalization is the $g_A$ quenching in nuclear Gamow-Teller transitions~\cite{gA,multifarious-gA}.  It has been shown that the hidden scale invariance emerges {\it precociously} in nuclear medium via strong nuclear correlations to lead to  an effective $g_A$, say $g_A^{\rm ss}\to 1$, approaching what is referred to as ``dilaton limit fixed point." The Lagrangian as given in (\ref{eq:LOLL}) is in the leading order in scale-chiral expansion~\cite{CT,LMR}.  In the sector where $\eta^\prime$ plays no role -- or negligible role -- and the dilaton field is ignored, the HLS Lagrangian is {\it gauge-equivalent} to nonlinear sigma model, therefore  one can do a systematic chiral-perturbation calculation similar to the standard $\chi$PT~\cite{HY:PR}.  The treatment of many-body systems resorted to in \cite{gA} is in  Landau Femi-liquid fixed point theory (FLFP) with the Lagrangian (\ref{bsHLS}) (given below) that figures in \cite{MR-review} amounts to doing nuclear higher-loop renormalization calculations in Wilsonian-RG approach on Fermi surface and what leads to $g_A^{\rm eff}\to 1$ aptly captures the restoring of scale symmetry hidden at the tree level to pseudo-conformal symmetry at the quantum level. 

Now going to high density beyond the normal nuclear matter, an important issue of the EoS of massive compact stars is the role of dilaton and indispensably the scale symmetry in QCD.  The story of scale symmetry in gauge theories has a long history dating from 1960's and it remains still a highly controversial issue in particle physics going beyond the Standard Model (BSM). Here we will confine ourselves to QCD for $N_f\leq 3$ for which we adhere to the notion that $f_0(500)$ is a ``genuine" dilaton being associated with hidden scale symmetry~\cite{crewther, CT}.  The distinctively characteristic feature of the ``genuine dilaton" scenario (GDS for short) is the presence of the IR fixed point signaling the scale invariance at which both scale and chiral symmetries are in  the NG mode admitting massive particles, such as nucleons, vector mesons etc.\footnote{It may be that this notion of the GDS is not widely accepted in the particle physics community working on BSM~ \cite{appelquist}. It seems however consistent with the notion of hidden quantum scale invariance~\cite{QSI}}  In our approach to dense matter it turns out,  as recalled below, that the GDS is consistent with the general structure of scale symmetry that manifests as an emerging symmetry from nuclear correlations at what we call ``dialton-limit fixed point (DLFP)" in dense matter.

\section{Baryonic matter without $\eta^\prime$}
We first consider baryonic matter where $U_A(1)$ anomaly does not figure. In (\ref{eq:LOLL}), we set $\eta^\prime$ equal to zero or properly integrated out given its massiveness in nature. The property of dense matter described by the theory, G$n$EFT, is analyzed in some detail in \cite{MR-review}. How to address many-nucleon systems directly from the Lagrangian that contains meson fields only, that is in the class of skyrmion approach, has not been worked out in a way suitable for dense matter physics. Therefore a direct exploitation of the Lagrangian (\ref{eq:LOLL}) treated entirely in terms of skyrmions is not feasible at present for studying the properties of dense baryonic matter. However an astute way is to  map what are established to be ``robust"  topological properties of skyrmions obtained with (\ref{eq:LOLL}) to a density functional-type theory -- referred frequently to as ``DFT" in nuclear physics circles -- by introducing explicitly baryon fields, and suitably coupling them,  to (\ref{eq:LOLL}). The strategy is to capture as fully as feasible non-perturbative properties associated with the topological structure involved. One possible way of how this can be done is discussed in detail in \cite{MR-review}.  Here  we summarize what one finds in the mean-field approximation of G$n$EFT which corresponds to doing Landau fixed-point  theory. Going beyond the approximation can be formulated in what is known as ``$V_{lowk}$" renormalization-group (RG) approach and applied to compact stars in \cite{PKLMR,MR-review}.
\subsection{Dilaton limit fixed point (DLFP)}
\label{DLFP}
To exploit the mapping of topological inputs of the sHLS Lagrangian into a  mean-field approximation with G$n$EFT,  we add the nucleon coupling to the sHLS fields implementing both HLS and scale symmetry as
\be
{\mathcal L}_N
& = & \bar{N}(i\Slash{D} - \Phi m_N ) N
{}+g_A \bar{N}\Slash{\hat{\alpha}}_\perp\gamma_5 N\nonumber\\
& & {}+  \bar{N}\left(g_{V\rho}\Slash{\hat{\alpha}}_\parallel + g_{V0} {\Tr} [\Slash{\hat{\alpha}}_\parallel] \right)  N +\cdots \,, \label{nucleon}
\ee
with the covariant derivative $D_\mu = \partial_\mu - iV^\rho_\mu - iV^\omega_\mu$
and dimensionless parameters $g_A$, $g_{V\rho}$ and $g_{V0}\equiv \frac 12 (g_{V\omega}-g_{V\rho})$. The ellipsis stands for higher derivative terms that will not be taken into account in what follows. The Lagrangian concerned that we shall refer to as $bs$HLS is
\be
{\mathcal L}_{bsHLS}={\CL}_{sHLS} + {\mathcal L}_N.\label{bsHLS}
\ee
With the explicit presence of the baryon field, the role of the hWZ terms is relegated to the  baryon-field coupling to the vector and scalar fields that takes over the $\omega$ repulsion in dense baryonic matter.

We consider what happens when the density goes up and approaches the DLFP first considered by Beane and van Kolck~\cite{bira-dlfp} and apply it to the model we are considering in \cite{WGP-interplay,PKLMR}. To do this we assume that approaching the DLFP at high density amounts to going toward the IR fixed point \`a la CT~\cite{crewther,CT} described above where both chiral symmetry and scale symmetry are realized in the NG mode with the dilaton mass and pion mass going to  zero in the chiral limit.

Starting from the vacuum where chiral symmetry is realized nonlinearly, as density increases,  one would like to arrive, at some point near $n_0$, at the linear realization of chiral symmetry, say, in the form of the Gell-Mann-L\'evy (linear) sigma model~\cite{gell-mann} which qualitatively captures nuclear dynamics as the Walecka mean-field model does. This means transforming the nonlinear structure of sHLS that is the habitat of the skyrmion structure to a form more adapted to dense matter, namely the half-skyrmion structure developed for the EoS of massive compact stars in \cite{PKLMR,MR-review}. This feature of transformation is encoded in the hidden scale symmetry as pointed out by Yamawaki~\cite{yamawaki}. This point will be further elaborated on in Section \ref{conclusion}.

To see how the $bs$HLS Lagrangian behaves as density is increased, we follow Beane and van Kolck~\cite{bira-dlfp} and transform $bs$HLS to a linear basis, $\Sigma=\frac{f_\pi}{f_\chi}U\chi\propto \sigma^\prime +i\vec{\tau}\cdot{\vec{\pi}}^\prime$.  We interpret taking the limit {${\mathcal S} \equiv {\rm Tr}(\Sigma^\dagger \Sigma)\to 0$} as approaching the  DLFP. Now how to take the dilaton limit requires a special interpretation. In mapping the key information of topological structure of baryonic matter to G$n$EFT as explained in \cite{MR-review}, it is essential to interpret the limiting ${\mathcal S} \to 0$ in the same sense as in going from the skyrmion phase to the half-skyrmion phase at a density above that of normal nuclear matter. In going from skyrmions to half-skyrmions in skyrmion crystal simulation,  the quark condensate $\la\bar{q}q\ra$ is found to globally go to zero at a density denoted  $n_{1/2} > n_0$.\footnote{To give an idea, $n_{1/2}$ in massive compact stars comes in \cite{MR-review} at $\sim 3n_0$.}  The condensate however is non-zero locally, thereby supporting a chiral density wave in skyrmion crystal~\cite{CDW}. This seems to be the case in general as observed in various models~\cite{buballa}. As a consequence, the pion decay constant is non-zero, hence the state is in the NG mode. The same is true for the dilaton condensate with inhomogeneity in consistency with the GDS. This feature resembles the ``pseudogap" structure in condensed matter physics. As there the issue is subtle and highly controversial (see \cite{pseudogap} for a comprehensive discussion on this matter). In what follows we interpret the limiting ${\mathcal S} \to 0$ in this sense. The order parameters for the symmetries involved in medium up to the possible IR fixed point of \cite{crewther} are more complicated involving higher dimensional field operators~\cite{Harada-Sasaki}. Approaching the DLFP, the quantities involved  will be denoted by asterisk $\ast$ as $\la\chi\ra^\ast\propto f_\pi^\ast$ and $\la\chi\ra^\ast\propto f_\chi^\ast$ with $f_\pi^\ast \sim f_\chi^\ast\neq 0$. The matter in the half-skyrmion phase going toward the DLFP then has a resemblance to the pseudo-gap phase with fractional skyrmions present in SYK superconductivity~\cite{gorskyetal}.

One finds that in the limit ${\mathcal S}\to 0$  there develop singularities in the thermodynamic potential. Imposing that there be  no  singular terms in that limit gives what we identify as DLFP ``constraints"~\cite{PKLMR}
\be
f_\pi \to f_\chi\neq  0, \ g_A\to g_{V\rho}\to 1.\label{dlfp-constraints}
\ee
Furthermore since the $\rho$-meson coupling to the nucleon is given by
\be
g_{\rho NN} = g_\rho (g_{V\rho}-1),
\ee
one sees that the $\rho$ meson decouples -- independently of the ``vector manifestation (VM)" with $g_\rho\to 0$~\cite{HY:PR} --  from the nucleons as the DLFP is approached. On the other hand, the $\omega$-NN coupling $g_{\omega NN} = g_\omega (g_{V\omega}-1)$ remains nonzero  for $g_\omega\neq 0$ because $g_{V\omega}-1 \not\to 0$ in the DLFP limit. This has been verified at one-loop order in the RGE~\cite{WGP-interplay}. In fact the EoS at high density relevant to massive compact stars requires this for the stability of the matter. It is well-known that there is a delicate balance between the dilaton condensate  which enters in the dilaton mass $m_\chi^\ast$ and the $\omega$-NN coupling $g_{\omega NN}$. In fact this balance is the well-known story of the roles of the scalar attraction and vector repulsion in nuclear physics at normal nuclear matter density. It becomes more acute at higher densities.
\begin{figure}[ht]
\begin{center}
\includegraphics[width=9.1cm]{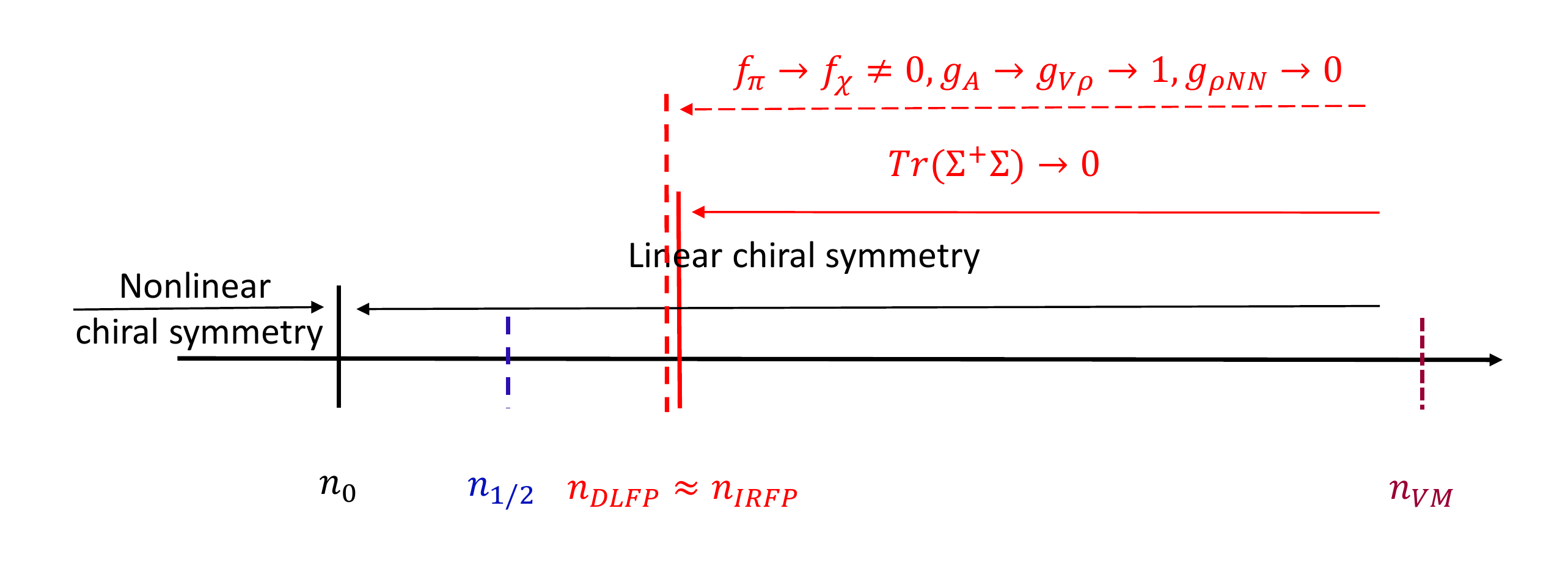}
\caption{The proposed schematic phase structure for density regimes:  $n_0$ stands for equilibrium nuclear matter density, $n_{1/2}$ for onset density of half-skyrmions, $n_{\rm DLFP}$ for dilaton limit fixed point, $n_{\rm IRFP}$ for IR fixed point and $n_{\rm VM}$ for vector manifestation fixed point.
}
\label{DensityClas}
\end{center}
\end{figure}

The broad phase structure involved is depicted in Fig.~\ref{DensityClas}. Apart from the nuclear matter equilibrium density $n_0$ and the topology change density $n_{1/2}$, other densities are not precisely pinned down. What's given in the review \cite{MR-review} does not represent precise values, hence Fig.~\ref{DensityClas} should be taken at best highly schematic.
%
\subsection{Interplay between $g_{\omega NN}$ and $\la\chi\ra^\ast$}

\begin{figure}[h]
\begin{center}
\includegraphics[width=6cm]{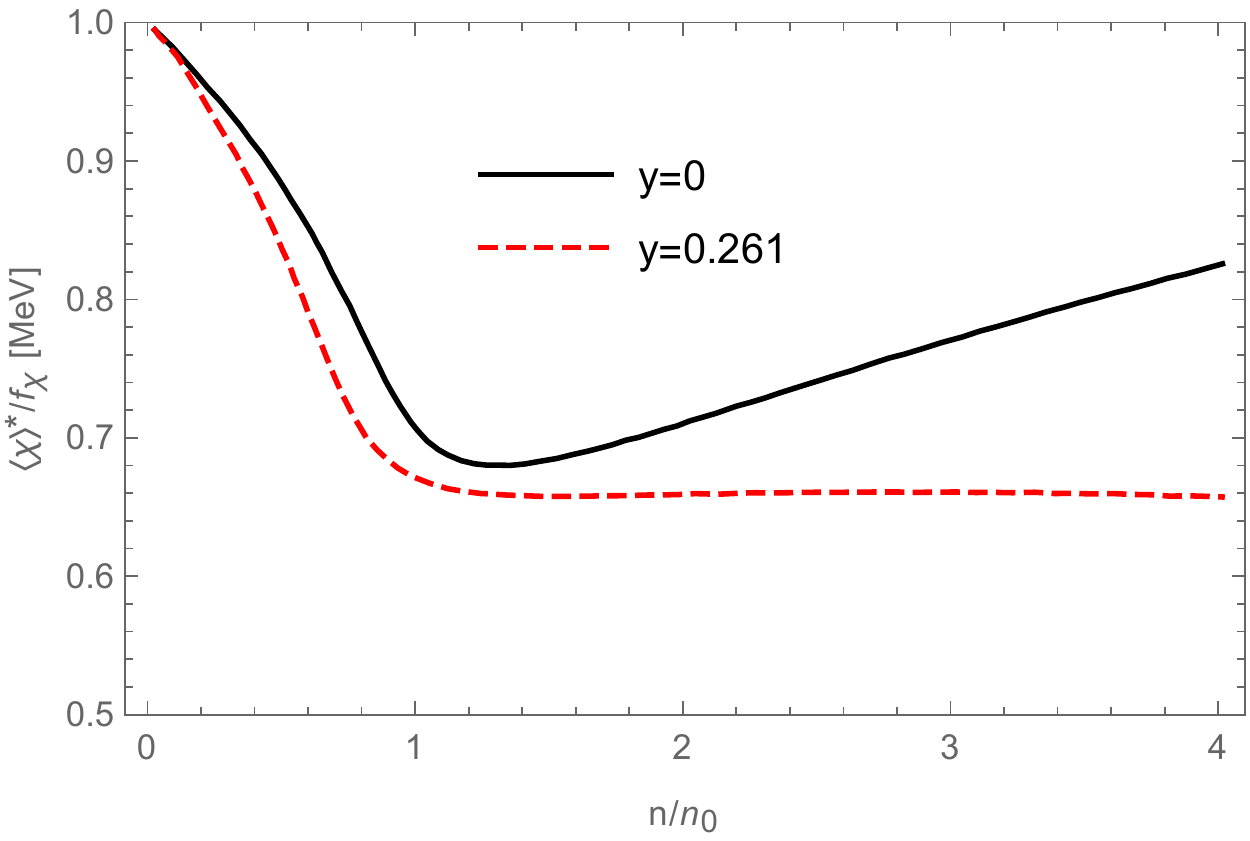}
\caption{The ratio $\la\chi\ra^*/\la\chi\ra_0$  where $\la\chi\ra^\ast\propto f_\chi^\ast$  as a function of
density $n$ for varying ``induced density dependence" (DD$_{induced}$) -- distinct from IDD (intrinsic density dependence) inherited from QCD -- of $g_{V\omega}^\ast$  which is  parameterized as  $g_{V\omega}^\ast -1=(g_{V\omega}-1)(1+y \frac{n}{n_0})^{-1}$.  The density at which the ratio $\la\chi\ra^*/\la\chi\ra_0$ becomes constant is not given by the theory but comes out to be $\sim 3 n_0$ in compact-star phenomenology. This density can be identified with $n_{1/2}$, the density at which skyrmion matter transitions to half-skyrmion matter.
}
\label{mass_coupling}
\end{center}
\end{figure}

The nucleon in-medium mass is connected to the $\omega$-nucleon coupling by the
equations of motion for $\chi$ and $\omega$ and the in-medium property
of the $\chi$ condensate, $\la\chi\ra^\ast$, or more appropriately the in-medium dilaton decay constant $f_\chi^\ast$ which controls the in-medium mass of the dilaton, hence
the nucleon mass, at high density. This means that up to the DLFP, the effective nucleon mass will remain constant proportional to the dilaton condensate $\la\chi\ra^\ast$. This is seen in Fig. \ref{mass_coupling}. This $\la\chi\ra^\ast$ comes out to be equal to the scale-chiral invariant mass of the nucleon  $m_0$ that figures in the parity-doubling model for the nucleon~\cite{sasakietal}. This then suggests that $m_0$ {\it can show up} signaling the presence at a higher density of the DLFP  through {\it strong nuclear correlations} even if it is not explicit in the QCD Lagrangian.

We claim that this is in accord with the GDS (``genuine dilaton" scenario) with the nucleon mass remaining massive together with the non-zero dilaton decay constant.

The remarkable interplay between the dilaton condensate $\la\chi\ra^\ast$ and the $\omega$-NN coupling has an important impact on the EoS for density $n\gsim n_{1/2}$ at which $\la\chi\ra^\ast$ flattens in density\footnote{The flattening to a density-independent constant of  $\la\chi\ra^\ast/f_\chi$ at $n_{1/2}$ arising from an intricate interplay between $g_{\omega NN}$ and $\la\chi\ra^\ast$ in Fig.~\ref{mass_coupling} is related to that of $\la\bar{q}q\ra^\ast/f_\pi$ in a skyrmion-crystal simulation of HLS~\cite{ohetal}. It is not obvious how to correctly implement the dilaton field in the crystal simulation, so the relation between the dilaton and quark condensates does not seem to come out correctly on skyrmion crystals. But in the genuine dilaton scenario incorporated in G$n$EFT, we believe they should be tightly related -- as we argued -- as density approaches the IR fixed point density.}. As noted above, the induced density dependence for the $\rho$-NN coupling $\propto (g_{V\rho}-1)$ drops rapidly such that the $\rho$  decouples from nucleons at the DLFP whereas ($g_{V\omega}-1$) remains non-zero. How this impacts on nuclear tensor forces and consequently on the symmetry energy $E_{sym}$ deserves to be investigated in nuclear structure. Furthermore the vector manifestation leads to the gauge coupling $g_\rho\to 0$~\cite{HY:PR} whereas that for $\omega$ coupling $g_\omega$ drops only slightly. The delicate balance between the attraction due to the scalar (dilaton) exchange and the repulsion due to the $\omega$ exchange plays a crucial role for the EoS for $n\gsim n_{1/2}$  of massive neutron stars~\cite{PKLMR}.

\subsection{The trace anomaly and pseudo-conformal symmetry}
A striking consequence of the interplay between the $g_{\omega NN}$ coupling and the condensate $\la\chi\ra^\ast$ at $n\gsim n_{1/2}$ in the G$n$EFT framework, not shared by other models in the literature, is the precocious emergence of hidden scale symmetry in nuclear interactions. The details are involved but the phenomenon can be clearly seen in the mean-field approximation with the $bs$HLS Lagrangian (\ref{bsHLS}).

The vacuum expectation value of the trace of the energy-momentum tensor $\theta_\mu^\mu$ is given by
\be
\la\theta_\mu^\mu\ra=4V(\la\chi\ra) -\la\chi\ra \left.\left(\frac{\del V(\chi)}{\del \chi}\right)\right|_{\chi=\la\chi\ra^\ast} +\cdots\label{MF-TEMT}
\ee
where the ellipsis stands for chiral symmetry breaking (quark mass) terms. Now if one ignores the quark mass terms, then given that the $\la\chi\ra^\ast$ which should be identified with the dilaton decay constant is independent of density~\cite{MR-review}, we have
\be
\frac{\del \la\theta_\mu^\mu (n)\ra }{\del n}=\frac{\del\epsilon (n)}{\del n} \big(1-3v_s^2/c^2\big)=0. \label{TEMT0}
\ee
One expects that $\frac{\del\epsilon (n)}{\del n}\neq 0$ and hence, within the range of density where (\ref{TEMT0}) holds, say, $\sim (3-7) n_0$,  we arrive at what is commonly associated with the ``conformal sound speed"
\be
v_s^2/c^2=1/3.\label{conformal-vs}
\ee
{\it Since however the trace of the energy-momentum tensor is not zero at the density involved which is far from asymptotic, it should be more appropriately called ``pseudo-conformal" velocity.}

This prediction made in the mean field for a neutron star of mass $M\simeq 2 M_\odot$ has been confirmed -- modulo of course quark-mass terms --- in going beyond the mean-field approximation using the $V_{lowk}$RG approach~\cite{PKLMR,MR-PCM}. Needless to say the quark mass terms could affect this result bringing in possible deviation from (\ref{conformal-vs}), but it seems reasonable to assume that the corrections cannot be significant.
\section{Baryonic matter with $\eta^\prime$}\label{withetaprime}
So far baryonic matter without the $\eta^\prime$ degree of freedom is treated as density increases toward the DLFP. The baryons involved there are skyrmions for $N_f=2$. It has been assumed that the $U_A(1)$ anomaly plays no role at high density for compact-star physics.

However there are at least three reasons why the $\eta^\prime$ degree of freedom cannot be ignored in nuclear dynamics. First, it is known that the $U_A(1)$ anomaly plays a crucial role for the color-charge conservation in the CCP~\cite{coloranormaly} and consequently for the flavor-singlet axial-vector coupling constant of the proton $g_A^0 \ll 1$~\cite{gA0}. Second,  the $\eta^\prime$, though massive at low density, may become lighter and become relevant at high density. Third, it has been suggested that the FQH droplet structure of $N_f=1$ baryon~\cite{zohar} can be unified in scale-symmetric HLS theory with the skyrmion structure of $N_f\geq 2$ baryons~\cite{karasik}. The third, while giving a possible solution to the dichotomy problem raised in Section \ref{problem}, could influence the EoS at high density.

We now discuss how to approach quantum Hall droplets from skyrmions.


\subsection{From sHLS to the $\eta^\prime$ ring}

In following Karasik's arguments~\cite{karasik,karasik2}, we  take the sHLS Lagrangian (\ref{eq:LOLL}) and focus on the terms that involve the $\eta^\prime$ field in that Lagrangian. {\it In doing this manipulation, the role of the conformal compensator present to provide (Seiberg-)duality plays the crucial role in contrast to what's done in \cite{karasik,karasik2} where the the role of dilaton effects is doing double trace which is ignored.} With the baryons generated as solitons in sHLS,  the parameters of the Lagrangian contain  the ``intrinsic density dependence" (IDD) inherited from QCD at the matching between EFT and QCD.  First we ask what one should  ``dial" in the parameters of G$n$EFT -- in the spirit of the strategy used -- to access compact stars so that the system approaches  the $\eta^\prime$ sheet.  Next we ask whether high baryon density supplied by gravity makes the $\eta^\prime$ ring ``visible."

Suppose we increase density beyond $n_{1/2}$. Recalling from what we have learned in the mean-field result with (\ref{bsHLS}) (with the nucleon explicitly included), it seems reasonable to assume the $\rho$ decouples first at some density above $n_{1/2}$ before reaching the DLFP. Since the gauge coupling $g_\rho$ goes to zero in approaching  the vector manifestation fixed point  $n_{VM}$ (say, $n\gsim 25 n_0$)~\cite{PKLMR}), the mass $m_\rho\sim f_\pi g_\rho$ goes to zero independently of whether or not $f_\pi$ goes to zero and the $\rho$ decouples from the pions. The Lagrangian ${\CL}_{sHLS}$ (\ref{eq:LOLL}) will then reduce to what was written down by Karasik~\cite{karasik,karasik2}
\be
{\CL}_{sHLS} &=& \frac 12 (\del_\mu\chi)^2 +V(\chi) + \frac{1}{2} {\Phi}^2 (\del_\mu \eta^\prime)^2 \nonumber\\
 &&{} - \frac{1}{4} (\omega_{\mu\nu})^2 + \frac{1}{2}m_\omega^2\Phi^2 \omega_\mu \omega^\mu \nonumber\\
 & &{} - \kappa \frac{N_c}{8\pi^2} \epsilon^{\mu\nu\alpha\beta} \omega_\mu\del_\nu\omega_\alpha\del_\beta\eta^\prime
 +\cdots\label{sHLSp},
\ee
where we have written $f = f_\eta$ and $f_{\sigma\rho}^2 = af_\pi^2$. The coefficient $\kappa$ is related to the coefficient $c_3$ in the hWZ term (\ref{hWZterm}) for the HLS dynamically generated, appropriate for low density.  We assume that  it will become quantized  as in \cite{karasik} when the $\eta^\prime$ ring is ``probed" at some high density as explained below.\footnote{In \cite{karasik2}, it has been argued that the well-known vector dominance in the presence of electroweak fields leads to the same constraint.}  In this operation, we assumed that the limit  ${\mathcal S}\to 0$ is equivalent to having,  as in the crystal simulation,  both the dilaton and chiral condensates -- space averaged globally -- go to zero while they locally support density waves {\it with their decay constants remaining  nonzero}. As density increases further beyond the DLFP,  the condensate will vanish locally and the kinetic energy term of $\eta^\prime$ field gets killed and the {$\omega$ mass $m_\omega\propto \la\chi\ra$} goes to zero. {Indeed, it is explicitly shown in \cite{CrystalHLS} in the crystal approach that at a density $n \gg n_{1/2}$ the phase becomes homogeneous -- without density waves -- so that $f_\pi \propto f_\chi \propto \langle \chi \rangle =0$. We interpreted this phase as deconfined since both chiral and scale symmetries are restored.}
We will be left with the massless $\chi$ and $\omega$ fields and  the  $\omega\eta^\prime$ coupling  coming from ${\CL}_3$ in the  hWZ term, (\ref{hWZterm}). 

{\it It is here that our approach to scale symmetry \`a la GD~\cite{crewther,QSI} brings the role of the sliding dilaton condensate at high density into  contact via the Seiberg-type duality with the $\eta^\prime$ ring structure.}

The last term of (\ref{sHLSp}) can be written as
 \be
{\CL}_{CS\eta^\prime}=-\kappa\frac{N_c}{4\pi} J_{\mu\nu\alpha} \omega^\mu\del^\nu\omega^\alpha. \label{CS-eta}
\ee
with the topological $U(1)$ 2-form symmetry current
\be
J_{\mu\nu\alpha}=\frac{1}{2\pi}\epsilon_{\mu\nu\alpha\beta} \del^\beta \eta^\prime.
\ee
Now a highly pertinent observation is that the charged objects under these symmetries get metamorphosed to infinitely extended sheets that interpolate from $\eta^\prime=0$ on one side to $\eta^\prime=2\pi$ on the other~\cite{zohar,karasik}, involving a sheet $\eta^\prime=\pi$.  The current is conserved because $\eta^\prime$ in the space of $\eta^\prime$ configuration is a circle and $\pi_1(S^1)={\cal Z}$.  The Lagrangian (\ref{CS-eta}) corresponds to the CS field identified with the $\omega$ field coupling to the baryon charge. The CS field is a topologically non-trivial gauge field and hence gauge invariance requires that the $\kappa$ be quantized $\kappa=1$.\footnote{ This argument holds if we assume that the matter is in the topological phase where HLS is Seiberg-dual to QCD. And also with the vector dominance~\cite{karasik2}. Karasik identifies the hWZ term with constrained coefficients as ``hidden WZ" term in contrast to the ``homogenous WZ term"~\cite{HY:PR} relevant at lower density. Note that going from ``homogeneous hWZ" to ``hidden hWZ" by density would require the unhiding of hidden scale invariance~\cite{crewther,QSI}.} This follows from the presumed duality between the gluon fields in QCD and the HLS fields in EFT~\cite{Y}.  Now going beyond the DLFP  as the system is brought toward the putative density  at which $\la\chi\ra\to 0$ and $\la\bar{q}q\ra\to 0$, $f_\pi\sim f_\chi\to 0$ and $m_\chi\to 0$, $m_\omega\to 0$ etc,  then one winds up with a quantum Hall baryon with $B=1$ and $J=N_c/2$~\cite{zohar}. This essentially rephrases Karasik's argument in terms of the G$n$EFT to  arrive at the $N_f=1$ baryon from the $N_f=2$ baryons.

In what's described above, we have assumed that the $\rho$ field decouples first before reaching the DLFP as indicated in Section \ref{DLFP}. This is what seems to take place in compact-star matter studied in \cite{PKLMR,MR-review}. Instead of $U(1)$ CS field theory, however, one can generalize the discussion to nonabelian CS field theory from the sHLS Lagrangian (\ref{eq:LOLL}). The Lagrangian (\ref{CS-eta}) is modified to~\cite{Y}
\be
{\CL}^\prime_{CS\eta^\prime}=-\kappa \frac{N_c}{4\pi}  J_{\mu\nu\alpha} {\rm Tr} \left(V^\mu\del^\nu V^\alpha +\frac 23 V^\mu V^\nu V^\alpha\right)
\label{eq:NonAbCS}
\ee
where $V_\mu=\frac 12 (\tau\cdot \rho_\mu +\omega_\mu)$,  assuming $U(2)$ symmetry,  is restored at the high density concerned.  We now identify the source of the baryon number as $B=(N_c/N_f) {\Tr} V$ and differentiating the action $\int {\CL}^\prime_{CS\eta^\prime}$ with respect to $B$, we get, with $\kappa=1$, the baryon density
\be
\rho_B=\frac{1}{4\pi^2}\epsilon^{ijk} {\Tr}(\del_i V_j)\del_k\eta^\prime +\cdots.
\ee
With the configuration $\eta^\prime =0$ at $x_3= - \infty$ and $\eta^\prime=2\pi$ at $x_3=\infty$,  the baryon number is gotten $B=1$ by integrating over $x_3$.
In \cite{Y}, an interpretation of this phenomenon is  made in terms of an anyon excitation of the quantum Hall droplet with baryon number $B=1/N_c$ leading to a quark described as a soliton made of hadrons, what one might interpret as a novel manifestation of hadron-quark duality.

{\it This exposes the $\eta^\prime$ ring in the $N_f=2$ setting.} This observation is relevant to the possible decay of the $\eta^\prime$ ring to a pionic sheet described in the next subsection.

Above we have seen that in some density regime, one arrives at a CS theory coupled to a baryon-charge one object that could be identified with the $\eta^\prime$ ring. This is done, we suggest, by what amounts to going beyond the DLFP in the G$n$EFT Lagrangian\footnote{For doing this more realistically, it may be necessary to include higher-lying vectors and scalars as in holographic models~\cite{Y}. This is beyond our scheme so we won't go into the matter further.}.
What's interesting is to view the process  in terms of the $N_f=2$ skyrmion given by the sHLS Lagrangian, namely, how the hedgehog ansatz in the background of the $\eta^\prime$ field is ``deformed" as density goes up. It seems plausible, as suggested in \cite{karasik}, that high density first impacts on the EoS such that
\be
\pi_1=\pi_2=0, \ V_\mu^1=V_\mu^2=0, V_\mu^3=\omega_\mu\label{V3}
\ee
and then distorts the hedgehog configuration to
\be
(\sigma +i\pi_3)/\sqrt{\sigma^2+\pi_3^2} =e^{i\eta^\prime/2f_\eta}.\label{etaprime}
\ee
This suggests that while at low density  $\eta^\prime$ in {$U=e^{i\eta^\prime/f_\eta} e^{i\tau\cdot\pi/f}$} present in the $\eta^\prime$ ring plays no significant role, except perhaps, giving an  $O(N_c^0)$ correction to the $\Delta-N$ mass difference which could not be significant,  the $\eta^\prime$ ring  becomes important as density increases.

\subsection{Going from the $\eta^\prime$ ring to pionic sheet}
We consider the density regime where the $\rho$ mesons are decoupled from the nucleons  and the $\eta^\prime$ ring is unstable, so decays to skyrmions.
Noting that the $\eta^\prime$ ring, i.e., {${\CL}_{CS\eta^\prime}$}, is embedded in the full  hWZ term, we should look at the hWZ term (\ref{hWZterm}). Following \cite{karasik}, we write (in the unitary gauge $\xi_R=\xi_L^\dagger=\xi$)
\be
{\CL}_{hWZ} &=& \frac{N_c}{24\pi^2} \epsilon^{\mu\nu\rho\sigma}g_\omega\omega_\mu \nonumber\\
& &  \times {\Tr} \Bigg[\left(\frac{3}{8}\kappa_1\right)2 \del_\nu\xi \xi^\dagger\del_\rho\xi\xi^\dagger\del_\sigma\xi\xi^\dagger\nonumber\\
& &\qquad\quad\; + \left(\frac{1}{2}\kappa_2\right) 3iV_\nu (\del_\rho\xi\del_\sigma\xi^\dagger -\del_\rho\xi^\dagger\del_\sigma\xi) \nonumber\\
& &\qquad\quad\; + \left(\frac{1}{2}\kappa_3\right) 3i\del_\nu V_\rho (\del_\sigma\xi\xi^\dagger - \del_\sigma\xi^\dagger\xi)\Bigg]
\ee
where only the terms contributing to the $N_f = 2$ completion of the topological term ${\cal L}_{CS\eta^\prime}$ are retained. The coefficients $\kappa_i$s can be identified with $c_i$s  of the hWZ term (\ref{hWZterm})\footnote{The coefficients $c_i$ have been fixed in \cite{karasik2} by imposing vector dominance (for which the $c_4$ term corresponding to the photon field is included). As will be remarked in the last Section, VD does not work well in nuclear physics unless the infinite tower of vector mesons of holographic QCD is included but it would be interesting to study their impact on dense matter.}
\be
\kappa_1=c_1-c_2, \ \kappa_2=c_1+c_2, \kappa_3=\ c_3.
\ee
Under gauge transformation $\omega_\mu \to \omega_\mu - \frac{1}{g_\omega}\partial_\mu \lambda$, one has
\be
\delta S &=& \frac{N_c}{12\pi^2} \epsilon^{\mu\nu\rho\sigma}\partial_\mu \lambda{\Tr} \Bigg[\left(\frac{3}{8}\kappa_1\right) \del_\nu\xi \xi^\dagger\del_\rho\xi\xi^\dagger\del_\sigma\xi\xi^\dagger\Bigg]\nonumber\\
& & + \frac{iN_c}{8\pi^2} \epsilon^{\mu\nu\rho\sigma}\partial_\mu \lambda \del_\nu{\Tr} \Bigg[ \left(\frac{1}{2}\kappa_2\right) V_\rho (\del_\sigma\xi\xi^\dagger - \del_\sigma\xi^\dagger\xi)\Bigg].\nonumber\\
\ee
Then the gauge invariance yields the constraints
\be
\int d\phi \left[\left(\frac{1}{2}\kappa_2\right)\left(\omega_\phi + V_\phi^3\right) + \left(\frac{3}{8}\kappa_1\right)4\pi_1\partial_\phi \pi_2\right]\nonumber\\
 = const.
\ee
So that, on the world-sheet for the $N_f=1$ baryon, one has~\cite{karasik}
\be
\frac{1}{2}\kappa_2\int d\phi(\omega_\phi +V_\phi^3) =2\pi, \ \pi_{1,2}=0.\label{E}
\ee
The $\eta^\prime$ ring is thereby ``seen."\footnote{We note that the mixing of two $U(1)$s $(\omega_\phi + V_\phi)$ in this formula (and also the equality $V_3=\omega$ in (\ref{V3})) follows from the assumed duality of HLS~\cite{kitanoetal} mentioned in footnote 2.}

Now suppose the $\eta^\prime$  sheet structure, a background buried in the system of $N_f=2$ skyrmions, is unstable and could subsequently decay into skyrmions in a different sheet structure containing the isovector degrees of freedom
\be
\omega_\phi +V_\phi^3=0, \ \ \frac{3}{8}\kappa_1\int d\phi \pi_1 \del_\phi \pi_2=\frac{\pi}{2}.\label{decayed}
\ee
The question  is: What is the structure of the matter encoded in the condition $\frac{3}{8}\kappa_1\int d\phi \pi_1 \del_\phi \pi_2=\pi/2$ to which the $\eta^\prime$ decays?  Could this be a sort of droplets that can be described in a topological field theory, involving isovector degrees of freedom, e.g., the $\pi^{\pm}$, the $\rho$ vectors etc. as in the form of a nonabelian CS Lagrangian that seems to arise in the Cheshire Cat for $N_f=2$ baryons? We have no answer to this question.  Clearly isovector mesons  must figure. This has to do with the quantization of other coefficients than the one giving the $\eta^\prime$ ring.
\section{Ubiquitous sheet structure of baryonic matter}
While it is not clear how the background of the $\eta^\prime$ ring, perhaps insignificant in the  dynamics of strongly interacting many-nucleon matter at low density, affects the process of going toward the DLFP -- and beyond -- to ``expose" the $\eta^\prime$ ring structure, it seems to be fitting to speculate how the QH droplets structure could manifest in the sheet structure of dense matter as seen in the EoS of massive compact stars described with fair success in \cite{PKLMR} and reviewed in \cite{MR-review}.
\subsection{Crystal skyrmions}\label{crystal}
We return to the skyrmion crystal simulation on which the G$n$EFT for massive compact stars is anchored~\cite{PKLMR,MR-review}. As detailed there, the topological structure of the skyrmions simulated on crystal is translated into the parameters of the G$n$EFT Lagrangian, which is then treated in an RG-approach to many-nucleon interactions. The key role played in this procedure is the topological feature encoded in the skyrmion structure of hidden scale symmetry and local symmetry of sHLS. Notable there are the cusp in the symmetry energy of dense matter due to the ``heavy" degrees of freedom\footnote{An intriguing observation  is that the ``cusp" in the symmetry energy (e.g., Fig.4 in \cite{MR-review}) arises at the next-to-the leading order in $1/N_c$ in the rotational quantization of the skyrmion energy in the Skyrme model (with pions only) as found first in \cite{Esym-cusp}. This observation was discussed in some detail in \cite{PREX}. The cusp in the $\eta^\prime$ potential in the Veneziano-Witten Lagrangian for $\eta^\prime$ is also higher order in $1/N_c$. Now when  the HLS vectors are ``integrated in" as done in \cite{karasik}, the cusp is no longer present.  Similarly when the HLS vectors are present in $Gn$EFT, the cusp also disappears as seen Fig. 7 in \cite{MR-review}. The symmetry energy involves the nuclear tensor forces and the cusp may be indicative of possible singularity in the one-pion exchange tensor force when given in terms of higher pion derivative terms in $Sn$EFT~\cite{bira-tensor}. The singularity could very well be eliminated by the $\rho$ meson as in $Gn$EFT. It would be interesting if this observation had something to do  in $N_f\geq 2$ baryons similarly  to what the HLS fields do to the $\eta^\prime$ singularity for the $N_f=1$ pancake baryon. This could lead to a possible resolution of the dichotomy problem.},
the parity doubling in the baryon spectra, and a ``pseudo-gap" structure of the half-skyrmion phase. These properties encapsulated in the RG-approach with G$n$EFT led to the prediction of possible precocious emergence of scale symmetry in massive-star matter with the pseudo-conformal sound velocity of star $v_{pcs}^2/c^2\simeq 1/3$ at a density $n\gsim3n_0$.

Let us explore what this skyrmion crystal structure  suggests for a possible sheet structure of dense matter.

It is observed in molecular dynamics simulation of nuclear matter expected in neutron-star crust and core-collapse supernova at a density a packing fraction of $\sim 5/16$ of  nuclear saturation density $n_0 \sim 2\times 10^{14}$g/cm$^3$ that a system of ``sheets" of lasagne, among a variety of complex shapes of so-called ``nuclear pasta," could be formed and play a significant role in the EoS in low-density regime of compact star matter~\cite{pasta}. Involved here are standard nuclear interactions between neutrons and protons in addition to electromagnetic interactions. 

At higher densities, say, at densities $\sim (2-7) n_0$,  it is seen in skyrmion crystal simulations that a stack of lasagne sheets~\cite{PPV} or of tubes or spaghettis ~\cite{canfora} is energetically favored over the homogenous structure. Involved here are  fractionalized skyrmions, 1/2-baryon-charged for the former and $1/q$-charged for the latter with $q$ odd integer. Those fractionalized skyrmions can be considered ``dual" to (constituent) quarks in  the sense of baryon-quark duality in QCD. There is an indication that the sheet structure of stack of lasagnes could give a consistent density profiles of  even finite nuclei~\cite{PPV}. In fact there seems little doubt that an inhomogeneity is favored in dense matter at {\it non-asymptotic} densities~\cite{CDW,buballa}. Thus it could be considered robust.

The two phenomena at low and high densities involve basically different aspects of strong interactions, but there is a tantalizing hint  that something universal is in action in both cases. We are tempted to consider that topology is involved there.  This is particularly plausible at high density given that the ``pasta" structure, be that lasagne or tubes (or spaghetti), is found to be strikingly robust. Up to date, the analysis has been made with an ansatz for the pion field, i.e., the Atiyah-Manton ansatz for the lasagne sheet and a special ansatz allowing analytical treatment for the tubes. The robustness must have to do with the fact that what is crucially involved is the topology and it is the pion field that carries the topology. What is striking is the resulting structure does not seem to depend on the presence of other massive degrees of freedom such as the vector mesons or scalar \cite{canfora2,CrystalHLS}. There, adding infinite number of higher derivative terms  to the Skyrme Lagrangian is found not to modify the ansatz for the tubes. It is therefore highly likely that the same structure would arise from the presence of the hidden scale-local degrees of freedom manifested in different forms of sHLS.
\subsection{Density functional theory (DFT)}
Our sHLS Lagrangian could contain the unified descriptions of both $N_f=1$ droplet -- $\eta^\prime$ ring -- and $N_f=2$ skyrmions in an EoS, but  we have not been able to capture both in a unified way. That is, how are the infinite-hotel and the FQH structures combined in the EoS and whether and how does the latter structure figure in compact-star physics?

A significant recent development,  relevant conceptually to this matter,  is the work treating the fractional quantum Hall  phenomenon in the Kohn's functional density approach \`a la  Kohn-Sham~\cite{jain}. The key ingredient in this approach is the weakly interacting composite fermions (CF) formed as bound states of electrons and (even number of) quantum vortices. Treated in Kohn-Sham density functional theory one arrives at the FQH state that captures certain strongly-correlated electron interactions.  The merit of this approach is that it {\it maps}  the Kohn-Sham density functional, a microscopic description, to the CS Lagrangian, a coarse-grained macroscopic description, for the fractional quantum Hall effect.

Now the possible relevance of this development to our problem is as follows. First,  Kohn-Sham theory more or less underlies practically {\it all} nuclear EFTs employed with success in nuclear physics, as for instance, energy density functional approaches to nuclear structure. Second, our G$n$EFT approach belongs to this class of density-functional theories in the strong-interaction regime. Third, the successful working of the  G$n$EFT model backed by robust topology and implemented with intrinsic density dependence inherited from QCD could very well be attributed to the power of the (Kohn-Sham-type) density functional in baryonic matter at high density $n\gsim 3n_0$. These three observations combined suggest to approach the {\it dichotomy problem} in a way related to what was done  for FQHE.

The first indication that G$n$EFT anchored on the topology change could be capturing the weak CF structure of \cite{jain} in FQHE is seen in the nearly non-interacting quasiparticle behavior in the chiral field configuration $U$ in the half-skyrmion phase (see Fig.~8 in \cite{MR-review}).  This feature may be understood as follows. Due to hidden $U(1)$ gauge symmetry in the hedgehog configuration, the half-skyrmion carries a magnetic monopole associated with the  dual $\omega$~\cite{cho,nitta}.  The energy of the ``bare" monopole in the half-skyrmion diverges when separated, but the divergence is tamed by interactions in the skyrmion as a bound state of two half-skyrmions where the divergence is absent. In a way analogous to what happens in the Kohn-Sham theory of FQHE~\cite{jain}, there  could  intervene the gauge interactions between the skyrmions pierced by a pair of monopoles in sHLS-- as composite fermions -- possibly induced by the Berry phases due to the magnetic vortices. Thus it is possible that the topological structure of the FQH is buried  in the bound half-skyrmion structure at high density.  A possible avenue to the problem is to formulate the EoS in terms of a stack of ordered coupled sheets of CS droplets. We hope to work this out in the future work.
\subsection{Hadron-quark continuity a.k.a. duality}
In the dilaton limit where the constraints (\ref{dlfp-constraints}) set in, there are NG excitations and the nucleon mass is $O(m_0)$ with $m_0\simeq y m_N$, $y\sim (0.6-0.9)$. In compact stars treated in \cite{PKLMR}, the core of the massive stars with $M\sim 2M_\odot$ has density $\sim (6-7) n_0$. A natural question one raises is whether the core of the star contains ``deconfined" quarks either co-existing with or without baryons. In the framework of \cite{MR-review}, the constituents of the core are fractional-baryon-charged quasiparticles. They are neither baryons nor quarks. The fractional-charged phase arises without order-parameter change and hence considered evolving continuously from baryonic phase with a certain topology change. In certain models having domain walls, those fractional-charged objects can be ``deconfined" on the domain wall~\cite{domainwall}. If the sheets in the skyrmion matter discussed above are domain walls, then it is possible that the fractional-charged objects are ``deconfined" on the sheets in the sense discussed in \cite{domainwall}.

There are two significant issues raised here.

One is the possible observation of an evidence for ``quarks" in the core of massive neutron stars~\cite{evidence}. Very recently, combining the astrophysical observations and theoretical {\it ab initio} calculations, Annal {\it et al.} concluded that inside the maximally massive stars there could very well be a quark core consisting of ``deconfined quarks"~\cite{evidence}.  Their analysis is based on the observation that in the core of the massive stars, the sound velocity approaches the conformal limit $v_s/c\to 1/\sqrt{3}$ and the polytropic index takes the value $\gamma < 1.75$, 1.75 being the value close to the minimal one obtained in the hadronic model.  It has been found~\cite{vs-core}  that in the pseudo-conformal structure of  our G$n$EFT, the sound velocity becomes {\it nearly} conformal $v_{pcs}/c\sim 1/\sqrt{3}$ and the $\gamma$ goes near 1 at $n \gsim 3 n_0$. Thus at least at the maximum density relevant for $\sim 2 M_\odot$ stars, what could be interpreted as ``deconfined quarks" can be more appropriately fractionally charged quasipartlcles. Are these ``deconfined" objects on domain walls as in \cite{domainwall} or confined two half-skyrmions as mentioned in Section \ref{crystal}? 

The other is what is referred to as ``baryon-quark continuity" in \cite{MR-review} in the domain of density relevant to compact-star phenomenology. This is not in the domain of density relevant to the color-flavor locking which is to take place at asymptotic density~\cite{wilczek}. It seems more appropriate to say that  the gauge degrees of freedom we are dealing with should be considered as ``dual" to the gluons in QCD~\cite{Y}.
\subsection{Hadron-quark continuity or deconfinement}
It has recently been argued that the hadron-quark continuity in the sense of \cite{wilczek} is ruled out on the basis of the existence of a nonlocal order parameter involving a  (color-)vortex holonomy~\cite{CFLruledout}. But such a ``theorem," perhaps holding at asymptotic density, could very well be irrelevant {\it even} at the maximum possible density observable in nature, whatever the maximum mass of the star stable against gravitational collapse might be. The argument of \cite{CFLruledout}  cannot rule out the baryon-quark duality argument given in \cite{MR-review} and in this paper which is far below asymptotic density. The presence of the scale-chiral-invariant nucleon mass $m_0$ testifies for this assertion.

\subsection{Emergence of hidden scale symmetry in nuclear matter}
The hadron-quark duality for which the hidden scale symmetry -- together with the HLS -- figures crucially in our discussion of resolving the dichotomy, we argue,  leaves a trail of other observables where its effect has impacts on. One prominent case is the long-standing mystery described above of the ``quenched" axial-vector coupling constant $g_A$ in nuclear medium reviewed in the note that accompanies this note~\cite{multifarious-gA}.

We recall as the density approaches the DLFP density, the constraints (\ref{dlfp-constraints}) require that the effective $g_A\to 1$. Surprisingly, as has been recently shown~\cite{gA},  the effective $g_A^{\rm ss}$ in Gamow-Teller transitions (most accurately measurable in doubly-magic nuclei) -- which is denoted  as $g_A^{\rm Landau}$ in \cite{gA} -- is predicted to be ``quenched" in the presence of emerging scale invariance from $g_A=1.27$ in free space to $g_A^{\rm ss}\approx 1$  at a much lower density, say, in finite nuclei.

Now here is an intriguing possibility. While  the precocious onset of the conformal sound speed $v_s^2/c^2=1/3$ in massive stars at density $n\gsim 3n_0$ is a signal for an emergence of an {\it albeit} approximate scale symmetry, there has been up to date no evident indication of the pseudo-conformal structure at low density $n\lsim 3n_0$. As suggested in \cite{WS-MPL}, it is an appealing possibility  that $g_A^{\rm eff}\approx 1$ in nuclei (modulo possible, as yet unknown, $\beta^\prime$ corrections) and $g_A=1$ at the DLFP is the continuity of the ``emergent" scale symmetry reflecting hidden quantum scale invariance between low and high densities.

Regardless of whether the hidden nature of scale symmetry is appropriate for the ``genuine dilaton" of Crewther~\cite{crewther}  or quantum scale invariance~\cite{QSI} or the dilaton in the conformal window~\cite{appelquist},  scale symmetry is intrinsically hidden. This point has been clearly illustrated in Yamawaki's argument~\cite{yamawaki}.  Yamawaki starts with the $SU(2)_L\times SU(2)_R$ linear sigma model with two parameters which corresponds to the Standard Model Higgs Lagrangian, makes a series of field re-parameterizations and writes the SM Higgs model in two terms, a scale invariant term and a potential term which breaks scale symmetry which depends on one dimensionless parameter $\lambda$. By dialing $\lambda\to \infty$, he gets the nonlinear sigma model with the scale symmetry breaking shoved into the NG boson field kinetic energy term, and by dialing $\lambda\to 0$ he gets scale-invariant theory going toward the conformal window. This suggests that one can think of what's happening in baryonic matter as dialing the parameter $\lambda$ in terms of nuclear dynamics. For compact-star physics, it's the density that does the dialing. 

\section{Comments and further remarks}\label{conclusion}

The principal proposition of this article is that the effective low-energy Lagrangian sHLS that incorporates hidden scale and local symmetries containing, in addition to the (octet) pions,  the $\eta^\prime$ degree of freedom could contain  both $N_f=1$ baryons and $N_f=2$ baryons through hidden scale and local symmetries dual to the gluons in agreement with Karasik. Our argument is admittedly far from rigorous. What is highly non-trivial is that the G$n$EFT could contain both the topological  structure of quantum Hall baryons and that of skyrmion baryons.  How to write the ansatz for the former as one does for the latter is unclear, but it should be feasible to do so and would allow one to see how the former comes in into the latter to resolve the dichotomy problem.

We are uncovering an interesting role that could  be played by the scale symmetry with its dilaton and the hidden local symmetry with the vector mesons dual to the gluons in ``unifying" the two different topological sheet structures. The analysis made in the G$n$EFT framework based on sHLS indicates that in the density regime relevant to massive compact stars, the chiral condensate and dilaton condensate go proportional to each other in the NG mode. In going beyond the regime of massive compact stars, we find the DLFP approaching, if not coinciding with,  the IR fixed point with $f_\pi = f_\chi$.  How and where the density regime for the IR fixed point  is approached cannot at present be elucidated.

It should be stressed again that the objective in this paper  is basically different from the fundamental issue of the role of topology and hidden symmetries in QCD and technicolor QCD. Firstly, what is evident is that what takes place in the dense compact-star regime  is drastically different from the presently favored scenario invoking the ``conformal window" in the domain of BSM with $N_f\sim 8$ typically characterized by the ratio~\cite{appelquist} $f_\pi^2/f_\chi^2 \sim 0.01$. This  is in stark contrast with   $f_\pi^2/f_\chi^2 \sim 1$ which seems to be encoded in the pseudo-conformal structure in dense nuclear systems. Secondly,  zeroing-in on the deep IR regime associated with the $\eta^\prime$ singularity involved in the domain-wall topological structure of baryons as argued in  \cite{karasik,karasik2} uncovers the $\omega$ (a.k.a. Chern-Simons) mass going to zero as the fermion (``quark") mass $m\to \infty$. This contrasts with how the $\eta^\prime$ ring structure is possibly ``exposed" in dense nuclear processes, as argued in this note, at high density as the $\omega$ mass is to go to zero with the dilaton mass $m_\chi\propto \la\chi\ra$ going to zero. This could be  explainable in terms of a  (Seiberg-type) duality between the gluons (in the topological phase)  and the HLS mesons (in the Higgs phase). In the former the vector dominance is found to play a crucial role for the $N_f=1$ baryon structure~\cite{karasik2}  whereas in the latter the VD -- unless a high tower of vector mesons  is taken into account~\cite{holography} -- famously fails to work for the $N_f=2$ (i.e., nucleon) EM form factors. How to correlate or reconcile these two processes appears highly challenging.\footnote{{\bf Note Added}: After this note was completed, we noticed the paper by Kitano and Matsudo~\cite{kitano} on the structure of the fractional quantum Hall pancake where the role of the hidden local gauge fields is basically different from that of Karasik~\cite{karasik} with which our approach overlaps, with some caveats, when applied to high-density matter.  At first reading, the Kitano-Matsudo approach appears to make certain predictions that render the key premise of the HLS adopted in our work at variance with their approach, e.g., the vector manifestation (VM) fixed point at which the vector meson mass is to vanish,  a key ingredient, though indirect, in the pseudo-conformal structure of dense matter and the identification of HLS fields as Chern-Simons fields at high densities in the vicinity of phase transitions. In response to the Kitano/Marsudo's argument,  Karasik~\cite{karasik2} has given a counter-argument to the effect that there is no essential disagreement between the two. The issue on the vector manifestation~\cite{HY:PR}, important in our work~\cite{MR-review}, remains however unresolved.}
\subsection*{Acknowledgments}
The work of YLM was supported in part by the National Science Foundation of China (NSFC) under Grant No. 11875147 and 11475071.

 \vfil

\end{document}